\documentclass{article}
\usepackage{ijcai23}

\usepackage{times}
\usepackage{soul}
\usepackage{url}
\usepackage[hidelinks]{hyperref}
\usepackage[utf8]{inputenc}
\usepackage[small]{caption}
\usepackage{graphicx}
\usepackage{amsmath}
\usepackage{amsthm}
\usepackage{booktabs}
\usepackage{algorithm}
\usepackage{algorithmic}
\usepackage{amsfonts}
\usepackage[switch]{lineno}

\title{TACOformer: Token-channel compounded Cross Attention for Multimodal Emotion Recognition}

\author{
     {Xinda Li}
    \affiliations 
     {Beihang University}
    \emails
    {richard519915@gmail.com}
}

\begin{document}
\maketitle

\begin{abstract}
    Recently, emotion recognition based on physiological signals has emerged as a field with intensive research. The utilization of multi-modal, multi-channel physiological signals has significantly improved the performance of emotion recognition systems, due to their complementarity. However, effectively integrating emotion-related semantic information from different modalities and capturing inter-modal dependencies remains a challenging issue. Many existing multimodal fusion methods ignore either token-to-token or channel-to-channel correlations of multichannel signals from different modalities, which limits the classification capability of the models to some extent. In this paper, we propose a comprehensive perspective of multimodal fusion that integrates channel-level and token-level cross-modal interactions. Specifically, we introduce a unified cross attention module called Token-chAnnel COmpound (TACO) Cross Attention to perform multimodal fusion, which simultaneously models channel-level and token-level dependencies between modalities. Additionally, we propose a 2D position encoding method to preserve information about the spatial distribution of EEG signal channels, then we use two transformer encoders ahead of the fusion module to capture long-term temporal dependencies from the EEG signal and the peripheral physiological signal, respectively. Subject-independent experiments on emotional dataset DEAP and Dreamer demonstrate that the proposed model achieves state-of-the-art performance.
\end{abstract}

\begin{figure}[ht]
  \includegraphics[width=\linewidth]{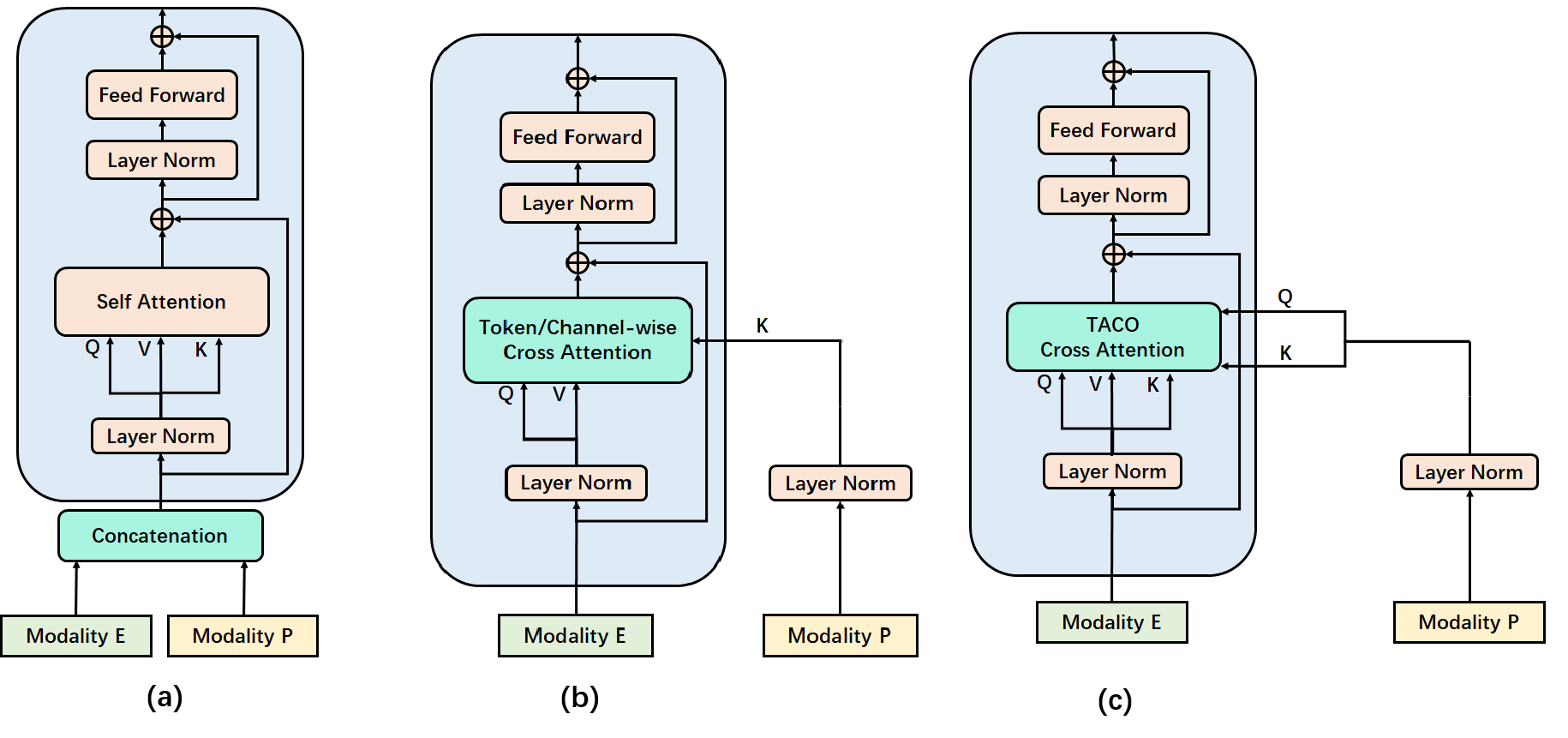}
  \caption{Comparison of the QKV attention-based multimodal fusion method: (a) Concatenation and self-attention (b) Token-wise (TCA) or channel-wise (CCA) cross attention (c) The proposed token-channel compounded (TACO) cross attention.}
  
  \label{fig:teaser}
\end{figure}

\section{Introduction}

Emotion is critical in human’s daily activities and interactions with the real world, affecting our cognition and behavior. The valence-arousal model characterizes emotion as a two-dimensional continuous space in which pleasantness or unpleasantness is described as valence, and excitement or calmness is described as arousal. Accurate emotion recognition can contribute to diverse fields like mental health monitoring, customized teaching programs, and security. Recent advancements in human-computer interaction technology have enabled the acquisition of multi-channel physiological signals, including Electroencephalography (EEG) and peripheral psychological signals (PPS) such as Electrocardiogram (ECG), Electromyogram (EMG), Electrooculogram (EOG), and Galvanic Skin Response (GSR). Due to their objectivity in reflecting emotional states compared to other approaches such as video, text, and audio, much research focuses on developing emotion recognition models based on physiological signals. Moreover, physiological signals from different modalities can provide complementary information about psychological states, which can help overcome noise that may arise from individual modality alone and reduce the ambiguity resulted from a single modality. Therefore, integrating multimodal physiological signals can make more accurate and robust predictions of emotion statements than unimodal emotion recognition methods.
\par
Many deep learning models have been built for processing multi-channel time series. But fusing these signals from different modalities remains a critical issue for constructing an emotion recognition model. 
\par
There mainly exist three strategies, which are on feature level, decision level, and model level, as outlined in the literature \cite{atrey2010multimodal}. The feature level fusion approach involves concatenation of features from various modalities to obtain a combined representation. Nonetheless, when dealing with signals comprising of numerous channels which corresponds to location of signal acquisition, spatial information in a single modality could be disregarded. Furthermore,  temporal synchronization information from different modalities may also be overlooked.
\par
The decision-level fusion methods involve feature extraction from each modality by passing input signal through specific network, embeddings from different networks are concatenated to get a joint representation. then a decision function outputs  final results. This approach provides the flexibility to select the most suitable method for each modality. In \cite{s20164551}, Convolutional Neural Networks (CNNs) and Long Short-Term Memory (LSTM) networks were separately employed to process the Electroencephalography (EEG), Galvanic Skin Response (GSR), and Electrocardiogram (ECG) signals. The Majority Vote was subsequently utilized as the decision function to produce predictions from different networks. Nevertheless, this approach has limited capabilities at capturing feature correlations among modalities. Furthermore, designing a heterogeneous network architecture could be time and resource-intensive.
\par
For model-level fusion, a large amount of studies concentrate on implement inter-modal interactions to model correlations between modalities. For example, \cite{liu2021comparing} employed a Weighted Sum Fusion method, which employed Deep Canonical Correlation Analysis (DCCA) to obtain coordinated representations of multimodal embeddings.
\cite{10.1145/3343031.3350871} employed a multimodal residual LSTM network to recognize emotion statements. This network is capable of learning the correlation between the Electroencephalography (EEG) and other physiological signals by sharing weights across modalities in each LSTM layer. \cite{hu2021mmgcn} used Graph Convolution Networks (GCNs) to model intra-modal and inter-modal dependencies among the audio, video, and text modalities. Cross attention \cite{9414654,9791482}is a widely adopted multimodal fusion technique that provides an effective approach for learning correlations of different tokens between modalities, which can be defined as token-wise cross attention (TCA). Furthermore, as different channels correspond to different locations for signal acquisition, adaptively fusing sufficient channel-wise features across modalities is advantageous for capturing spatial information. As such, channel-wise cross attention (CCA) \cite{wang2022uctransnet} has been utilized to learn the correlation of different channels between two modalities. However, the aforementioned cross attention mechanisms are based on a single perspective at implementing multimodal fusion, either token-wise or channel-wise. Therefore, it is essential to find an effective way to conduct interaction between modalities based on a compounded perspective of channel-wise and token-wise. 
\par
In this paper, we propose a network for emotion recognition based on multimodal psychological data (EEG signals and peripheral psychological signals like EOG, EMG, and ECG signals). The proposed multimodal fusion module, namely Token-chAnnel COmpounded(TACO) cross attention, enables us to simultaneously capture long-term token-wise and channel-wise dependence between two modalities. Figure \ref{fig:teaser} illustrates TACO cross attention and other two attention mechanism-based fusion methods. With TACO cross attention as the fusion method, we build TACOformer network, which utilizes two separate transformer encoders \cite{vaswani2017attention} as temporal extraction module.
The main contributions of this paper are summarized as follows:
\begin{itemize}
\item
We suggest a new perspective of  multimodal fusion, i.e. the compound of token-wise and channel-wise cross attention, which can simultaneously capture long-term compounded dependency between two multi-channel modalities on temporal level and channel level.
\item
We propose a 2D position encoding method to preserve spatial information in the input sequence of two-dimensional images, which outperforms 1d position encoding methods. 
\item
Extensive subject-independent experiments are conducted on two benchmark datasets and the experimental results show that our network consistently outperforms other models at valence and arousal classification.
\end{itemize}

The rest of paper is organized as below: section 2 introduces preliminary. Section 3 presents the proposed methodology. Section 4 introduces datasets and data preprocess. Section 5 presents experiments and analysis. Section 6 concludes this paper.

\begin{figure*}
\includegraphics[width=\linewidth]{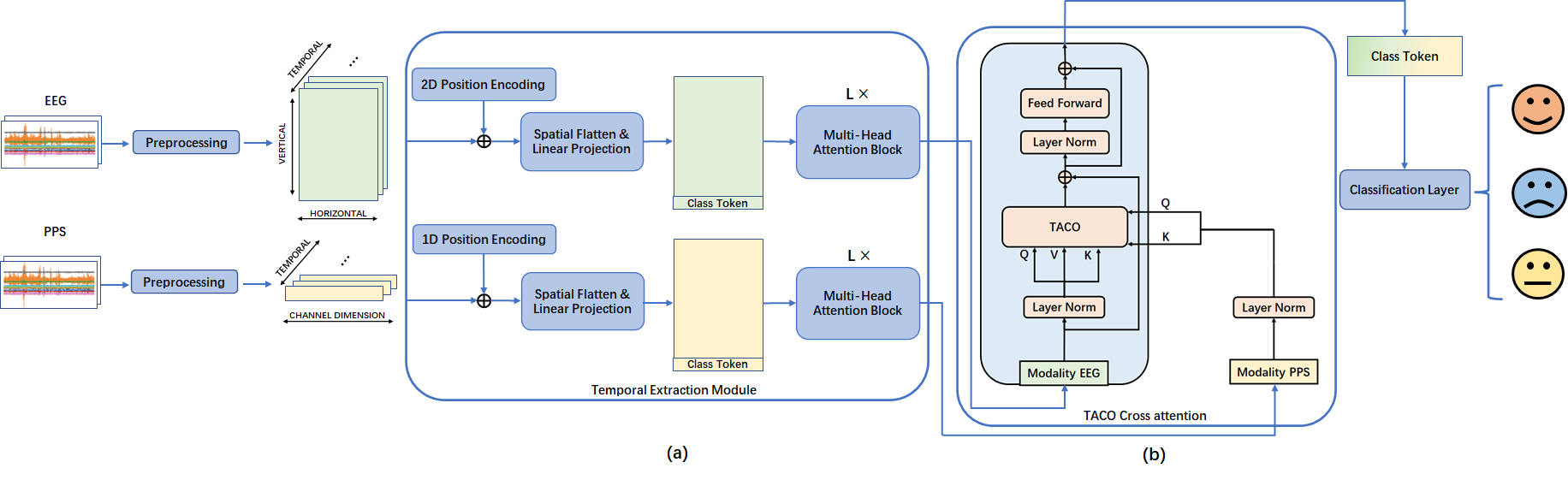}
\caption{(a) The whole process of emotion recognition. EEG and PPS signals are represented as 3D and 2D format input into network. The model consists of temporal extraction module (two independent transformer encoders), TACO cross attenion fusion module and classification layer. (b) structure of TACO corss-attention fusion, where we propose a cross attention mechanism based on compounded perspective of token-wise and channel-wise dependency.} \label{fig2}
\end{figure*}

\section{Preliminary}
In this paper, we define $X_e=(E_1,E_2,\cdots,E_T )\in \mathbb{R}^{N \times T} $ as an EEG signal sample containing $T$ time stamps, where $N$ is the number of electrodes, $E_t=(e_t^1,e_t^2,\cdots,e_t^N )\in \mathbb{R}^N $ denotes the EEG signal of $N$ electrode channels collected at the time stamp $t$. $E_t $ is then transformed into a 2D matrix $ E2D\in ^{H \times W}$  (see Section 4.2 Data preprocess), where $H$ and $W$ denote the height and width of the matrix. Further, we combine 2D matrices at $T$ timestamps then get a 3D spatial-temporal representation $X_{E}=(E2D_1,E2D_2,\cdots,E2D_T )\in \mathbb{R}^ {H\times W\times T}$ which is the input data of EEG signal. Similarly, PPS with $K$ channels is defined as  $X_P=(P_1,P_2,\cdots,P_T )\in \mathbb{R}^{K\times T}$, which contains $T$ timestamps. $P_t=(p_t^1,p_t^2,\cdots,p_t^K )\in \mathbb{R}^K$ denotes the peripheral physiological signal of $K$ channels at timestamp $t$.
The EEG emotion recognition problem is defined as: learning a mapping function which maps the input data that includes EEG and Peripheral physiological signals to the corresponding emotion:
\begin{equation}
Y_{class}=F(X_E,X_P)
\end{equation}
where $F$ denotes the mapping function, $Y_{class}$ denotes the predicted emotion class.

\section{Methodology}
\subsection{Model overview}
Figure \ref{fig2} illustrates the overall structure of TACOformer. It consists of temporal extraction module, TACO cross attention fusion, and classification layer.
After data preprocessing, EEG is represented as a temporal-spatial 3D matrix, PPS is represented as a 2D matrix. Two Multi-head attention transformers extract temporal features from EEG and PPS in the temporal extraction module. TACO cross attention models the compounded correlations between channel-to-channel and token-to-token of two modalities to obtain fused embedding. The classification layer utilizes aggregated information represented as classtoken and connects it with a linear layer to get classification output.

\subsection{2D Spatial Position Encoding}
Transformer is permutation-invariant when processing sequence, so it is necessary to implement position encoding to retain relative or absolute position information. However, 1D position encoding in \cite{vaswani2017attention} ignores spatial distribution of EEG channels. To tackle this problem, we generalize 1D sinusoidal position encoding to 2D  position encoding. Made by product of sines in vertical position and cosines in horizontal position, we get 2D position encoding $POS \in R^{T \times H \times W  }$, which is depicted as follows:

\begin{equation}
\resizebox{\linewidth}{!}{$
POS(t,x,y)= \begin{cases}
sin(t/ 10000^{x/ W})\cdot cos(t/ 10000^{y/ H})& \text{x=2i,y=2j}\\
cos(t/ 10000^{x/ W})\cdot cos(t/ 10000^{y/ H})& \text{x=2i+1,y=2j}\\
cos(t/ 10000^{x/ W})\cdot sin(t/ 10000^{y/ H})& \text{x=2i+1,y=2j+1}\\
sin(t/ 10000^{x/ W})\cdot sin(t/ 10000^{y/ H})& \text{x=2i,y=2j+1}\\
\end{cases}
$
}
\end{equation}
where $x$ and $y$ are the horizontal and vertical position, $t$ is the timestamp of token in sequence, $W$ and $H$ denote width and height of the 2D EEG matrix.

\subsection{Temporal Extraction Module}
Due to the capability at capturing long-range dependencies from sequential data of Transformer, we adopt the encoder part of Vanilla Transformer \cite{vaswani2017attention} to extract features from EEG and PPS. The key component of Vanilla Transformer is Multi-head attention mechanism. It is depicted as follow:
Given the input EEG vector $X_{E}\in \mathbb{R}^ {H\times W\times T}$ , where $T$ denotes the length of input sequence, $H$ and $W$ denote the height and width of EEG matrix. 
it is firstly transposed into shape as $T\times H\times W$ and 2D position encoding $POS$ is point-wise added on it. Spatial flattening and Linear projection transforms it into $Z_E \in \mathbb{R}^{T \times d}$. Following process in \cite{devlin2018bert}, we concatenate an learnable token $CLS \in \mathbb{R}^{1 \times d} $ as the first token in sequence and use $LayerNorm$ function to normalize embedding, which is depicted as follows:
\begin{equation}
    Z = LayerNorm(Concat(CLS, Z_E)) \in \mathbb{R}^{(T+1) \times d}
\end{equation}

Next,three embeddings Q(query),K(key),V(value) are generated from linear projections:
$Q=Z\times W^Q,K=Z\times W^K,V=Z\times W^V$ where $W^{Q},W^{K},W^{V}\in \mathbb{R}^{d\times d}$. The dot product of $Q$ and $K$ is used to derive a weight matrix, which is rescaled by multiplication with $1/{\sqrt{d}}$. The $Softmax$ function is applied along rows to generate the attention matrix:
\begin{equation}
M=Softmax(\frac{Q\times K^T}{\sqrt{d}})
\end{equation}
\begin{equation}
Attn=M\times V=Softmax(\frac{Q\times K^T}{\sqrt{d}})\times V
\end{equation}
where $n$ denotes the sequence length, $M_{i,j}$   are normalized alignment scores measuring the similarity between  tokens $z_i $and $z_j$.After residual connection and feed-forward network, the final output is defined as:
\begin{equation}
    Res= Z + Attn 
\end{equation}
\begin{equation}
     output= Res + FFN(LayerNorm(Res))
\end{equation}
where $FFN$ is a token-wise feed forward network, which is implemented by a linear layer with output's  dimension as $Res$.

For multi-head attention, input embedding $Z$ is splitted into $h$ smaller embeddings, then these embeddings are fed through separate linear projections to generate Query $Q_i$, Key $K_i$ and Value $V_i$ matrices for each attention head. Then $h$ head attention outputs are computed, concatenated and through a linear projection:
\begin{equation}
    Attn_{i} = Softmax(\frac{Q_i \times {K_i}^T}{\sqrt{d}})\times V_i\in \mathbb{R}^{n\times \frac{d}{h}} 
\end{equation}
\begin{equation}
    Attn_{MultiHead}= Concat(Attn_{Head1},\cdots,Attn_{headh})\times W^O
\end{equation}
where $W^O \in \mathbb{R}^{d \times d}$ is the output projection matrix.
As illustrated in Figure.2, embeddings of EEG and PPS are obtained from two independent transformer encoders.

\subsection{Token-channel compound Cross Attention }
\begin{figure}[ht]
\includegraphics[width=\linewidth]{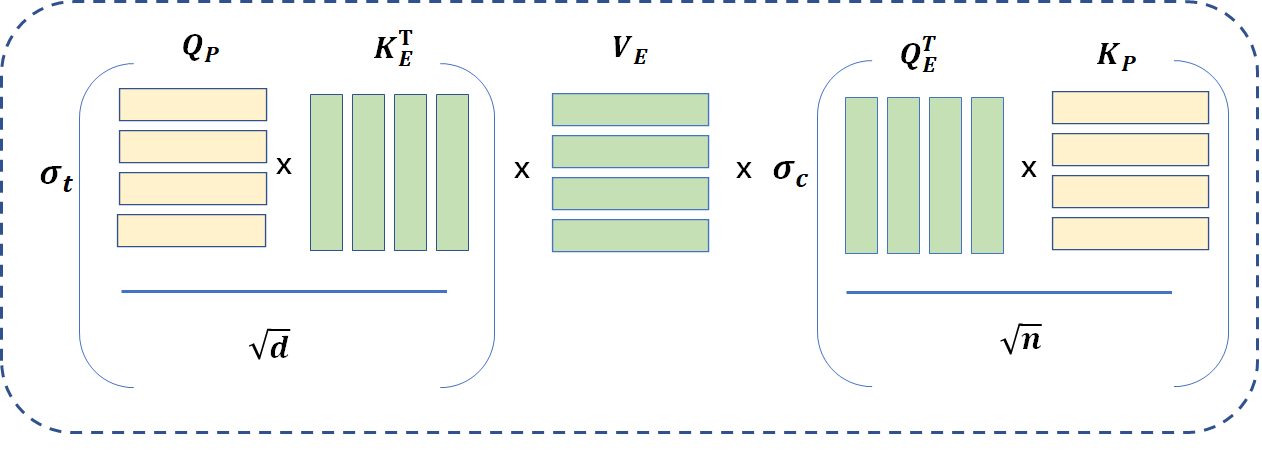}
\caption{Token-channel compounded cross attention mechanism. In which, channel-wise cross attention and token-wise cross attention are constructed by right part and left part of TACO, respectively.} \label{fig3}
\end{figure}

Token-wise cross attention (TCA) and channel-wise cross attention (CCA) can capture long-term dependencies between different tokens and between different channels of two modalities, respectively. However, on one hand, implementation of token-wise cross attention to conduct multimodal fusion could ignore channel-level dependencies, vice versa. On the other hand, simple combination like concatenation of the two cross attention matrix could cause data sparseness, which leads to a larger latent space and requires much more computing resource.
Therefore, we propose a novel cross attention mechanism to compound channel-wise and token-wise cross attention, it is depicted as follow:

\begin{equation}
TACO=\sigma_t (\frac{Q_P\times K_E^T}{\sqrt{d}})\times V_E \times \sigma_c (\frac{Q_E^T\times K_P}{\sqrt{n}})
\end{equation}
where $d$ denotes dimension of tokens in embedding space, $n$ denotes length of the token sequence. $Q_E\in \mathbb{R}^{n\times d}$ and $K_E \in R^{n\times d}$ are the Query and Key matrix from embedding of EEG $E$, $Q_P\in \mathbb{R}^{n\times d}$ and $K_P\in \mathbb{R}^{n\times d}$ are the Query and Key matrix from embedding of PPS $P$. $\sigma_t $ denotes $Softmax$ function along the rows, $\sigma_c$ denotes $Softmax$ function along the columns. It is illustrated in Figure \ref{fig3}.

In $TACO$ cross attention, token-wise and channel-wise correlations between EEG and PPS are modeled by the left and right part, which can correspond to $TCA$ and $CCA$ depicted as follows:  
\begin{equation}
TCA_{E\xrightarrow{}P}= \sigma (\frac{Q_P\times K_E^T}{\sqrt{d}})\times V_E 
\end{equation}

\begin{equation}
CCA_{P\xrightarrow{}E}=V_E\times \sigma (\frac{Q_E^T\times K_P}{\sqrt{n}})
\end{equation}
where $CCA_{P\xrightarrow{}E}\in \mathbb{R}^{n\times d}$ denotes the channel-wise cross attention matrix, which measures attention scores of channels of modality E in the perspective of channels in modality P, $TCA_{E\xrightarrow{}P}\in \mathbb{R}^{n\times d}$ denotes the token-wise cross attention matrix, which measures attention scores of tokens in modality P in the perspective of tokens in modality E.
\par 
The proposed TACO cross attention synthesizes the above two cross attention matrices, thus get a compounded result that measures correlations on token-level and channel-level simultaneously. Given the $TACO$ , final result is defined as:
\begin{equation}
    Res_{out} = TACO + E 
\end{equation}
\begin{equation}
    F_{out}=Res_{out}  + FFN(LayerNorm(Res_{out}))
\end{equation}
where $FFN$ is a token-wise feed forward network, which is implemented by a linear layer with output  dimension as $Res_{out}$.
The overall structure of TACO cross attention fusion module is illustrated in Figure \ref{fig2}.

\subsection{Classification Layer}
To Further integrate information from the fusion result $F_{out}$ which is composed of many tokens, we utilize linear layer with input as class token which is the first token of $F_{out}$to generate classification output
. Cross-entropy is used as loss function, which is defined as follows:
\begin{equation}
        L = -\sum_{c\in class set}{y_c log({P_{c}})}
\end{equation}
where $c$ denotes a class in class set, $y_c$ denotes the binary label (0/1) of class $c$ for certain input signal, $P_c$ denotes the predicted probability of class $c$ for the input.

\section{Experiments}
\subsection{Dataset and Preprocess}

\begin{figure}[ht]
\includegraphics[width=\linewidth]{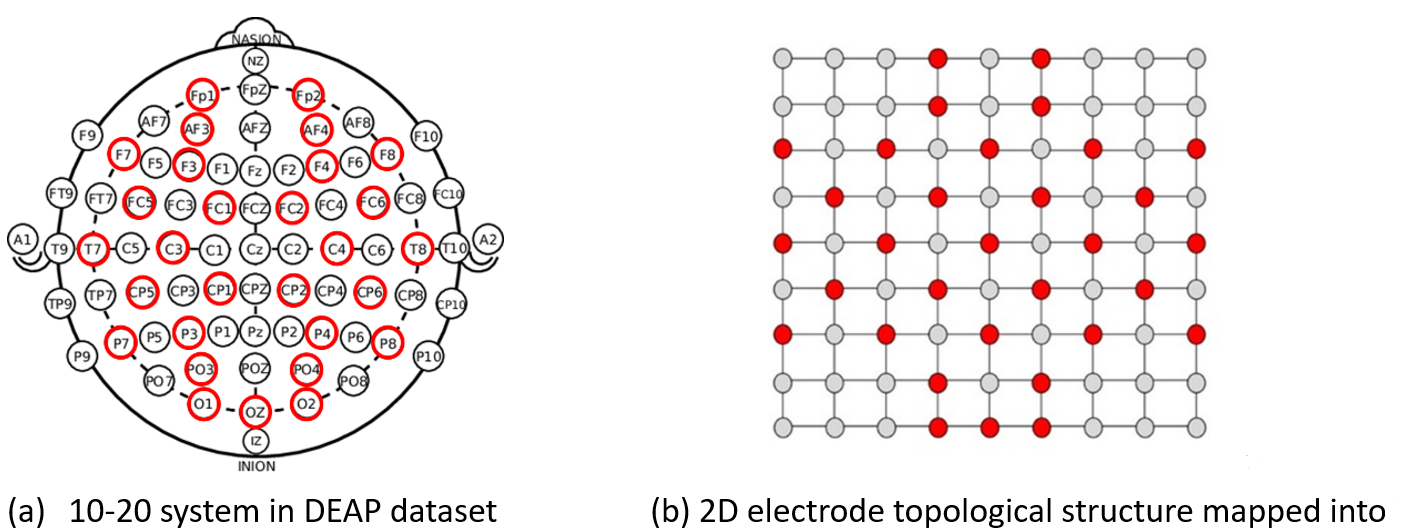}
\caption{(a) The international 10-20 system which describe the location of electrodes used for acquiring EEG signal, red nodes are used in DEAP dataset. (b) The 9$\times$9 matrix which represents 2D electrode topological structure.  } \label{fig4}
\end{figure}

\textbf{DEAP} \cite{koelstra2011deap} is a dataset focusing on analyzing human affective states based on multimodal signals. In this dataset, 32 subjects are examined, 40 videos each of which lasts 63s are selected as stimuli. While subjects are watching videos, EEG signals are acquired with electrodes placed as Figure \ref{fig4}, peripheral physiological signals like hEOGS, vEOGs, zEMGs, tEMGs are also recorded. The first 3s of 63s in each trial is acquired as baseline signal which is recorded when subjects are not under video stimuli. After watching each video, subjects are asked to conduct self-assessments of valence, arousal, dominance and liking where they rate them on a scale from 1 to 9. In our work, we use 5 as threshold and separate the assessment scores into a two-class label set.
\par
In data preprocessing of DEAP dataset, signal from one subject collected when watching one vidoe is a matrix with shape as 80$\times$ 8064, the row of which represents 32 channels of EEG and 8 channels of peripheral physiological signal, the column of which represents 3s' baseline and 60s' stimuli signal recorded in 128Hz. Baseline signal (a 40$\times$384 matrix) is cut into 3 segments (each segment is a 40$\times$128 matrix), the mean value (a 40$\times$128 matrix) of them is calculated. The mean value of baseline signal is subtracted from the stimuli signal along the time dimension to get data with shape as 40$\times$ 7680. For each timestamp, the 32-channel EEG signal is mapped into a 9$\times$9 matrix that represents the 2D electrode topological structure \cite{image_2dmap}, which is illustrated in Figure \ref{fig4}. After that, each 2D 9$\times$9 matrix is normalized with Z-score normalization. Finally, the processed data are cut into 60 segments, each of which includes 1s signal and has the shape of 9$\times$9$\times$128. According to \cite{9207420}, 1s is the most suitable time window length of signal for emotion recognition. Data size of EEG signals in DEAP after processing is 76800 instances (32 participants$\times$40 trials$\times$60s),the dimension of each instance is 9$\times$9$\times$128. The peripheral physiological signal of 8 channels is cut into 60 segments with 1s length in each modality, the data size of peripheral physiological signal in each modality after processing is 76800 instances, the dimension of each instance is 8$\times$128. We then randomly split total dataset with 20\% as test dataset and 80\% as training dataset.
\par
\textbf{Dreamer} \cite{katsigiannis2017dreamer} used audio-visual stimuli for affect elicitation from 23 subjects. Each subject  exposed to 18 different videos of variable length from 65 s–393 s duration where stimuli signals are recorded and a period of 61 s where subjects are in neutral emotion statement and baseline signals are recorded. Each subject was asked to label the valence and arousal values with the scale from 1 to 5 using Self-Assessment Manikins (SAM) after watching one video. Then we use 3 as threshold and separate the assessment scores into a two-class label set. 14 channels of EEG with AF3, F7, F3, FC5, T7, P7, O1, O2, P8, T8, FC6, F4, F8, AF4 channels were recorded using Emotiv-Epoc portable sensor with sampling rate as 128 Hz. ECG with two channels is recorded using the Shimmer sensor with sampling rate of 256 Hz.
\par
\begin{figure}[ht]
\includegraphics[width=\linewidth]{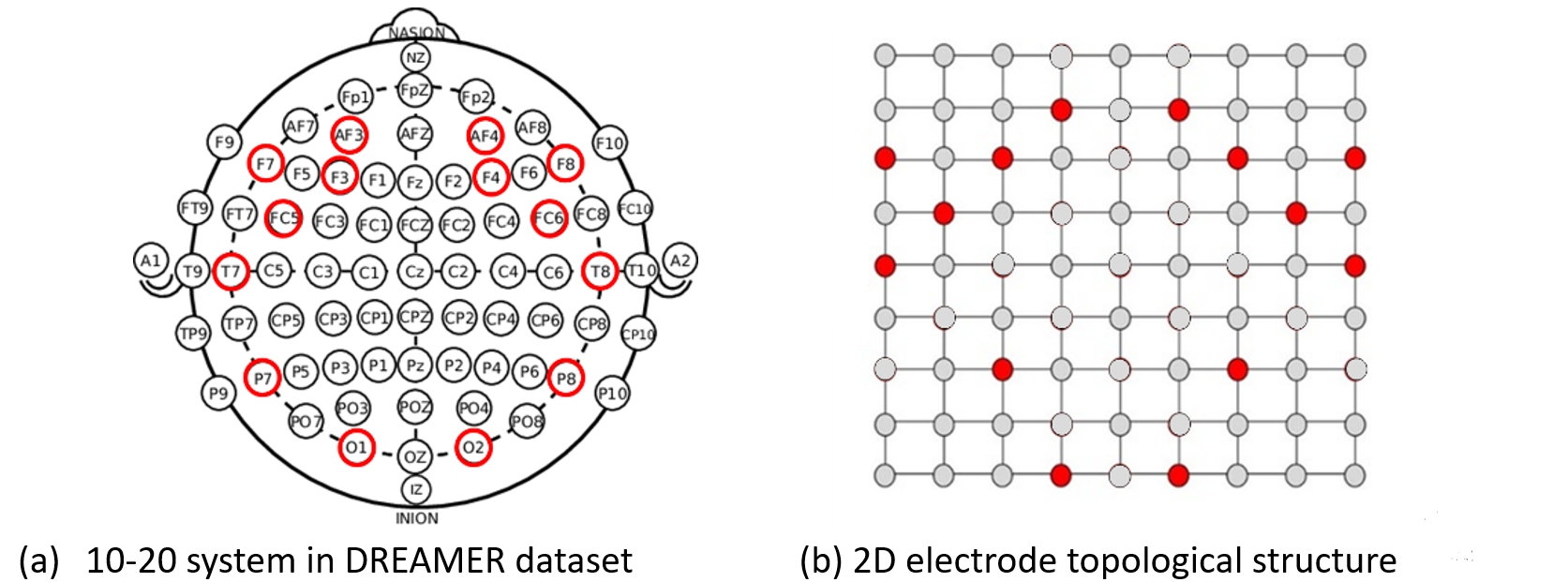}
\caption{(a) The international 10-20 system which describe the location of 14 electrodes used for acquiring EEG signal, red nodes are used in Dreamer dataset. (b) The 9$\times$9 matrix which represents 2D electrode topological structure.} \label{fig5}
\end{figure}

For Dreamer dataset that contains EEG and ECG signals, we first let stimuli signals subtract the mean value of all channels at each timestamp . Then, we utilize bandpass that ranges from 4-45Hz to filter the last 62s of stimuli signals. Moreover, we calculate a mean vector (14$\times$1) of 61s' baseline signal and subtract it from the filtered stimuli signal at each timestamp. After that, at each timestamp, the EEG signal of 14 channels is mapped into a 9$\times$9 matrix to represent 2D electrode topological structure as shown in Figure \ref{fig5} . Like DEAP dataset, each 9$\times$9 matrix is normalized with Z-score normalization and flatten. Finally, we get EEG signal dataset with 24840 instances (23 participants$\times$18 trials$\times$60s), the dimension of each instance is 9$\times$9$\times$128. And ECG signal after processing also contains 24840 instances, the dimension of each instance is 2$\times$ 128. We then randomly split total dataset with 20\% as test dataset and 80\% as training dataset. 

\subsection{Implementation details}
We train our model on single NVIDIA R5000X GPU with 24GB memory. The Adam optimizer is used to minimize the loss function. The learning rate is set to 0.001. Batch size is set as 120.
 For each experiment, we randomly shuffle the samples to get training dataset and testing dataset. The ratio of training set to test set is 4:1.

\subsection{Baseline models}
To verify the effectiveness of our proposed model, we compare it with the following models on the emotion recognition task.
\begin{itemize}

\item 
ECLGCNN\cite{yin2021eeg}: it extract entropy features from segmented data, then utilize LSTM and graph convolution neural network as classification model. 
\item
3DFR-DFCN\cite{9321553}: it extracts kurtosis, power, differential entropy from EEG signals in different frequency bands, then rearrange features to get a 3d representation. Then a fully convolution neural network is utilized to process features, a regularization term is used to minimize spectral norm of weight matrices.
\item
FLDNET\cite{9314183}: it utilizes LSTM to construct a hierarchical network to learn features at different granularity. Then 1d convolution and Softmax are used as decision layer to get output. 
\item
DCCA\cite{liu2021comparing}: it extracts features in frequency and time-frequency domain from EEG, ECG and PPS. Embeddings from two modalities are fused by weighted summation based on attention scalars.
\item
GA-MLP\cite{9588702}: it extracts power spectral density features from EEG as input to MLP model. Then genetic algorithm optimized the model configurations. 
\end{itemize}

\subsection{Result Analysis and Comparison}

We compare our model with other baseline models on DEAP and DREAMER dataset on subject-independent experiments. We adopt 5-fold cross validation experiment with random shuffling on original dataset to verify TACOformer. The left part of Table \ref{tab1} presents valence's and arousal's classification accuracy and standard deviation of five models for emotion recognition on DEAP dataset. The right part presents them on DREAMER dataset. The proposed TACOformer achieves the state-of-the-art performance on both datasets.
\begin{table}[htbp]

\begin{center}
\resizebox{0.47\textwidth}{0.06\textheight}{
\begin{tabular}{ccccc}
\hline
\textbf{Model}&\multicolumn{2}{c}{\textbf{DEAP}}&\multicolumn{2}{c}{\textbf{DREAMER}} \\

\textbf{ } & {\textbf{\textit{Valence ($\%$)}}}& \textbf{\textit{Arousal ($\%$)}}&  {\textbf{\textit{Valence ($\%$)}}}& \textbf{\textit{Arousal ($\%$)}} \\
\hline
ECLGCNN &84.81 &85.27 & - & - \\
3DFR-DFCN& 81.03&79.91 &82.49&75.97  \\
FLDNET&83.85 $\pm$11.34&78.82 $\pm$10.34&87.67 $\pm$10.02 &89.91 $\pm$12.51\\
DCCA&85.62&84.33&88.99&90.57\\
GA-MLP&88.28&90.63& - & -\\
\hline
\textbf{TACOformer} & \textbf{91.59 $\pm$ 0.51} & \textbf{92.02 $\pm$ 0.73}& \textbf{94.58 $\pm$ 4.73}&\textbf{94.03 $\pm$ 1.71}\\
\hline
\end{tabular}
}
\caption{\label{tab1}The comparison of the state-of-the-art models on the DEAP and DREAMER dataset}
\end{center}
\end{table}

\subsection{Ablation studies}
In this section, we verify the effectiveness and superiority of our proposed modules from two aspect: multimodal fusion module and spatial position encoding method.
\par
\textbf{Ablation on multimodal fusion modules.} We conducted experiments on multimodal fusion module. We set "2d position encoding+ temporal extraction module" as Backbone model and compare the performance of four models with different fusion module( concatenate, token-wise cross attention, channel-wise cross attention and TACO cross attention). Table \ref{tab2} demonstrates that “Backbone+ TACO cross attention” outperforms the other strategies, it indicates that both token-wise and channel-wise correlation are important, a compounded utilization of both methods is recommended.
\begin{table}[htbp]

\begin{center}
\resizebox{0.47\textwidth}{!}{
\begin{tabular}{ccc}
\hline
\textbf{Model}&\multicolumn{2}{c}{\textbf{DEAP}} \\

\textbf{ } & {\textbf{\textit{Valence ($\%$)}}}& \textbf{\textit{Arousal ($\%$)}}   \\
\hline
Backbone+Concat& 84.05& 86.18  \\
Backbone+TCA&87.51&87.40 \\
Backbone+CCA&88.79&89.62 \\
\hline
\textbf{Backbone+TACO(TACOformer)}& \textbf{91.59} & \textbf{92.72}\\ 

\hline
\end{tabular}

}
\caption{\label{tab2}Ablation study on multimodal fusion module}
\end{center}
\end{table}

\textbf{Ablation on position encoding method.} We also conducted experiments on position encoding module.
We compare our proposed 2d position encoding method with classic 1d position encoding method\cite{vaswani2017attention}. Table \ref{tab3} shows that the proposed 2d position encoding achieves better performance than 1d position encoding. It demonstrates that spatial distribution of multi-channel signal should be considered when building spatial-temporal network.
\begin{table}[htbp]

\begin{center}
\resizebox{0.47\textwidth}{!}{
\begin{tabular}{ccc}
\hline
\textbf{Model}&\multicolumn{2}{c}{\textbf{DEAP}} \\

\textbf{ } & {\textbf{\textit{Valence ($\%$)}}}& \textbf{\textit{Arousal ($\%$)}}   \\
\hline
1d position encoding& 89.34& 90.83  \\
\hline
\textbf{2d position encoding}& \textbf{91.59} & \textbf{92.72}\\
\hline

\end{tabular}
}
\caption{\label{tab3}Ablation study on position encoding method}
\end{center}
\end{table}

\subsection{Statistical Test}
To measure the significance of improvement of our proposed multimodal fusion method-- TACO cross attention, We conduct 5-fold cross validated paired t-test between frameworks that are based on TACO, CCA and TCA. Table \ref{tab5} shows the T-statistics and p-value of paired t-test, which are generated by comparing performance of TACO and TCA, TACO and CCA. Both p-values are less than 0.05, which demonstrates that there is a statistically significant difference in performance between "Backbone+TACO" and the other two models.

\begin{table*}[htb]
\begin{center}

\begin{tabular}{|c|c|c|c|c|}
\hline
\textbf{Model Pair}&\multicolumn{2}{c}{\textbf{Valence}} \vline&\multicolumn{2}{c}{\textbf{Arousal}}\vline\\
\hline
\textbf{} &\textbf{T-statistics} &\textbf{p-value}&\textbf{T-statistics} &\textbf{p-value} \\
\hline
\textbf{TACO and TCA}&14.848&0.0001 & 11.305& 0.0003 \\
\hline
\textbf{TACO and CCA}&13.872&0.0002& 3.154& 0.0343\\
\hline

\end{tabular}

\caption{\label{tab5}T-statistics and p-value of paired 5-fold cross validated t-test}
\end{center}
\end{table*}

\subsection{Visualization}
\begin{figure}[ht]
\includegraphics[width=\linewidth]{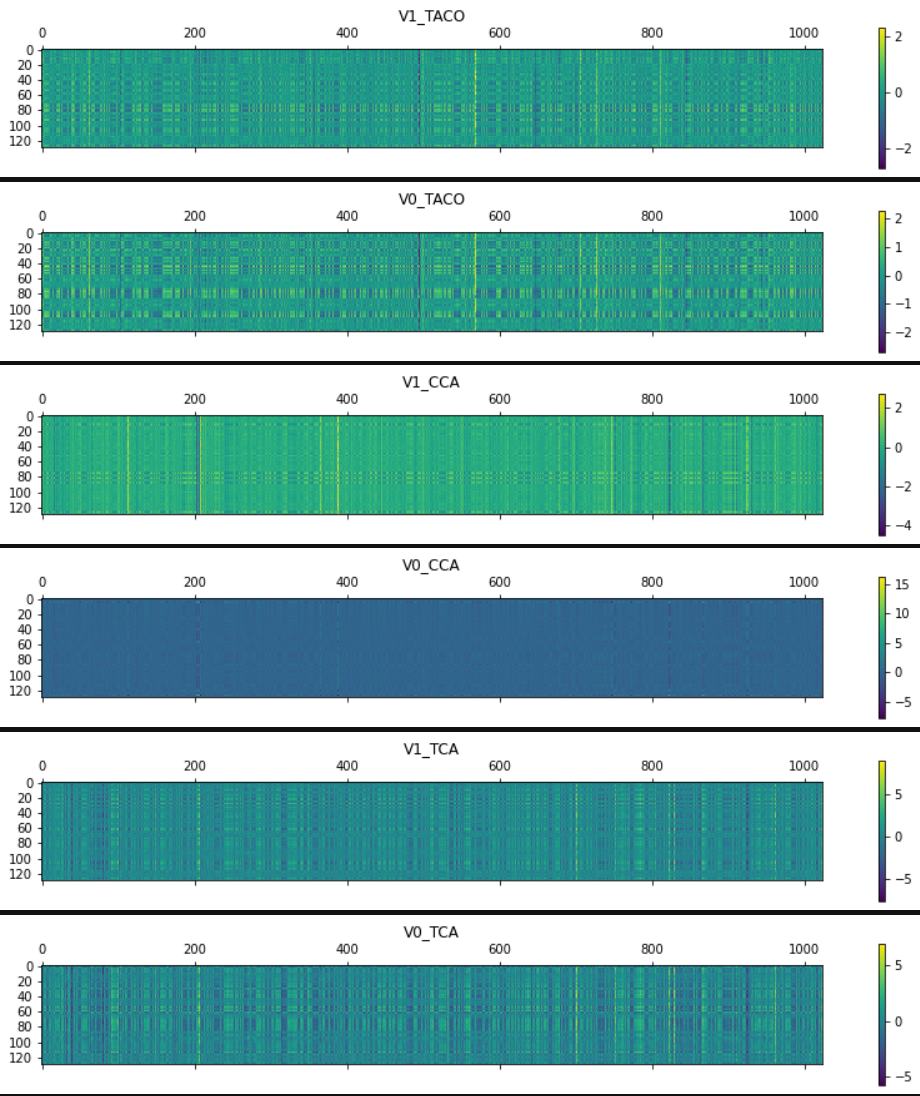}
\caption{Illustration of output of cross attention matrices- TACO, CCA and TCA. Head title with "V1" and "V0" denote high and low valence label,respectively.The horizontal axis represents the channel dimension, the vertical dimension represents the temporal dimension} \label{fig6}
\end{figure}

To perform a thorough evaluation on the proposed TACO cross attention, we visualize the cross attention fusion matrices in TACO, CCA, TCA. Figure \ref{fig6} illustrates the attention matrices from different cross attention methods, which are generated from input data corresponding to high and low valence.
The horizontal direction denotes the time dimension and vertical direction denotes the channel dimension.

Intuitively, we can see that TACO cross attention matrix varies larger on temporal dimension and channel dimension than CCA and TCA. Elements with large scores are distributed throughout different tokens and channels. But in CCA and TCA, same patterns repeats along channel axis and time axis, respectively. This indicates TACO can assign attention scores in the whole latent space.

\section{Conclusion}
In this paper, we propose a network for emotion recognition based on multimodal psychological signal. The proposed multimodal fusion module token-channel compound (TACO) cross attention takes a compounded perspective to conduct interactions between modalities. Meanwhile, the utilization of transformer as temporal extraction module effectively learns long-term dependencies and achieves high-accuracy classification. The subject-independent experiments on DEAP and DREAMER datasets demonstrate TACOformer achieves better performance than the other baselines. In addition, ablation studies show the effectiveness of the proposed modules including 2d position encoding and TACO cross attention. It is worth noting that the proposed TACO cross attention module is a general fusion method for multimodal multivariate time series, which can be further applied in other fields.

\bibliographystyle{named}
\bibliography{ijcai23}

\begin{thebibliography}{}

\bibitem[\protect\citeauthoryear{Atrey \bgroup \em et al.\egroup
  }{2010}]{atrey2010multimodal}
Pradeep~K Atrey, M~Anwar Hossain, Abdulmotaleb El~Saddik, and Mohan~S
  Kankanhalli.
\newblock Multimodal fusion for multimedia analysis: a survey.
\newblock {\em Multimedia systems}, 16:345--379, 2010.

\bibitem[\protect\citeauthoryear{Dar \bgroup \em et al.\egroup
  }{2020}]{s20164551}
Muhammad~Najam Dar, Muhammad~Usman Akram, Sajid~Gul Khawaja, and Amit~N.
  Pujari.
\newblock Cnn and lstm-based emotion charting using physiological signals.
\newblock {\em Sensors}, 20(16), 2020.

\bibitem[\protect\citeauthoryear{Devlin \bgroup \em et al.\egroup
  }{2018}]{devlin2018bert}
Jacob Devlin, Ming-Wei Chang, Kenton Lee, and Kristina Toutanova.
\newblock Bert: Pre-training of deep bidirectional transformers for language
  understanding.
\newblock {\em arXiv preprint arXiv:1810.04805}, 2018.

\bibitem[\protect\citeauthoryear{Hu \bgroup \em et al.\egroup
  }{2021}]{hu2021mmgcn}
Jingwen Hu, Yuchen Liu, Jinming Zhao, and Qin Jin.
\newblock Mmgcn: Multimodal fusion via deep graph convolution network for
  emotion recognition in conversation.
\newblock In {\em Proceedings of the 59th Annual Meeting of the Association for
  Computational Linguistics and the 11th International Joint Conference on
  Natural Language Processing (Volume 1: Long Papers)}, pages 5666--5675, 2021.

\bibitem[\protect\citeauthoryear{Katsigiannis and
  Ramzan}{2017}]{katsigiannis2017dreamer}
Stamos Katsigiannis and Naeem Ramzan.
\newblock Dreamer: A database for emotion recognition through eeg and ecg
  signals from wireless low-cost off-the-shelf devices.
\newblock {\em IEEE journal of biomedical and health informatics},
  22(1):98--107, 2017.

\bibitem[\protect\citeauthoryear{Koelstra \bgroup \em et al.\egroup
  }{2011}]{koelstra2011deap}
Sander Koelstra, Christian Muhl, Mohammad Soleymani, Jong-Seok Lee, Ashkan
  Yazdani, Touradj Ebrahimi, Thierry Pun, Anton Nijholt, and Ioannis Patras.
\newblock Deap: A database for emotion analysis; using physiological signals.
\newblock {\em IEEE transactions on affective computing}, 3(1):18--31, 2011.

\bibitem[\protect\citeauthoryear{Li \bgroup \em et al.\egroup
  }{2017}]{image_2dmap}
Youjun Li, Jiajin Huang, Haiyan Zhou, and Ning Zhong.
\newblock Human emotion recognition with electroencephalographic
  multidimensional features by hybrid deep neural networks.
\newblock {\em Applied Sciences}, 7:1060, 10 2017.

\bibitem[\protect\citeauthoryear{Li \bgroup \em et al.\egroup }{2021}]{9321553}
Dongdong Li, Bing Chai, Zhe Wang, Hai Yang, and Wenli Du.
\newblock Eeg emotion recognition based on 3-d feature representation and
  dilated fully convolutional networks.
\newblock {\em IEEE Transactions on Cognitive and Developmental Systems},
  13(4):885--897, 2021.

\bibitem[\protect\citeauthoryear{Liu \bgroup \em et al.\egroup
  }{2021}]{liu2021comparing}
Wei Liu, Jie-Lin Qiu, Wei-Long Zheng, and Bao-Liang Lu.
\newblock Comparing recognition performance and robustness of multimodal deep
  learning models for multimodal emotion recognition.
\newblock {\em IEEE Transactions on Cognitive and Developmental Systems},
  14(2):715--729, 2021.

\bibitem[\protect\citeauthoryear{Lu \bgroup \em et al.\egroup }{2022}]{9791482}
Houhong Lu, Yangyang Zhu, Ming Yin, Guofu Yin, and Luofeng Xie.
\newblock Multimodal fusion convolutional neural network with cross-attention
  mechanism for internal defect detection of magnetic tile.
\newblock {\em IEEE Access}, 10:60876--60886, 2022.

\bibitem[\protect\citeauthoryear{Ma \bgroup \em et al.\egroup
  }{2019}]{10.1145/3343031.3350871}
Jiaxin Ma, Hao Tang, Wei-Long Zheng, and Bao-Liang Lu.
\newblock Emotion recognition using multimodal residual lstm network.
\newblock In {\em Proceedings of the 27th ACM International Conference on
  Multimedia}, MM '19, page 176–183, New York, NY, USA, 2019. Association for
  Computing Machinery.

\bibitem[\protect\citeauthoryear{Marjit \bgroup \em et al.\egroup
  }{2021}]{9588702}
Shyam Marjit, Upasana Talukdar, and Shyamanta~M Hazarika.
\newblock Eeg-based emotion recognition using genetic algorithm optimized
  multi-layer perceptron.
\newblock In {\em 2021 International Symposium of Asian Control Association on
  Intelligent Robotics and Industrial Automation (IRIA)}, pages 304--309, 2021.

\bibitem[\protect\citeauthoryear{Sun \bgroup \em et al.\egroup
  }{2021}]{9414654}
Licai Sun, Bin Liu, Jianhua Tao, and Zheng Lian.
\newblock Multimodal cross- and self-attention network for speech emotion
  recognition.
\newblock In {\em ICASSP 2021 - 2021 IEEE International Conference on
  Acoustics, Speech and Signal Processing (ICASSP)}, pages 4275--4279, 2021.

\bibitem[\protect\citeauthoryear{Vaswani \bgroup \em et al.\egroup
  }{2017}]{vaswani2017attention}
Ashish Vaswani, Noam Shazeer, Niki Parmar, Jakob Uszkoreit, Llion Jones,
  Aidan~N Gomez, {\L}ukasz Kaiser, and Illia Polosukhin.
\newblock Attention is all you need.
\newblock {\em Advances in neural information processing systems}, 30, 2017.

\bibitem[\protect\citeauthoryear{Wang \bgroup \em et al.\egroup
  }{2021}]{9314183}
Zhe Wang, Tianhao Gu, Yiwen Zhu, Dongdong Li, Hai Yang, and Wenli Du.
\newblock Fldnet: Frame-level distilling neural network for eeg emotion
  recognition.
\newblock {\em IEEE Journal of Biomedical and Health Informatics},
  25(7):2533--2544, 2021.

\bibitem[\protect\citeauthoryear{Wang \bgroup \em et al.\egroup
  }{2022}]{wang2022uctransnet}
Haonan Wang, Peng Cao, Jiaqi Wang, and Osmar~R Zaiane.
\newblock Uctransnet: rethinking the skip connections in u-net from a
  channel-wise perspective with transformer.
\newblock In {\em Proceedings of the AAAI conference on artificial
  intelligence}, volume~36, pages 2441--2449, 2022.

\bibitem[\protect\citeauthoryear{Yin \bgroup \em et al.\egroup
  }{2021}]{yin2021eeg}
Yongqiang Yin, Xiangwei Zheng, Bin Hu, Yuang Zhang, and Xinchun Cui.
\newblock Eeg emotion recognition using fusion model of graph convolutional
  neural networks and lstm.
\newblock {\em Applied Soft Computing}, 100:106954, 2021.

\bibitem[\protect\citeauthoryear{Zhao \bgroup \em et al.\egroup
  }{2020}]{9207420}
Yuxuan Zhao, Jin Yang, Jinlong Lin, Dunshan Yu, and Xixin Cao.
\newblock A 3d convolutional neural network for emotion recognition based on
  eeg signals.
\newblock In {\em 2020 International Joint Conference on Neural Networks
  (IJCNN)}, pages 1--6, 2020.

\end{thebibliography}

\end{document}